\documentclass[conference]{sty/IEEEtran}

\usepackage{sty/cite}

\usepackage[dvips]{graphicx}

\usepackage[cmex10]{amsmath}

\usepackage{comment,color}

\begin{document}
%
\title{Loop-Back Interference Suppression \\ for OFDM Signals via Sampled-Data Control}



%
\author{\IEEEauthorblockN{Hampei Sasahara\IEEEauthorrefmark{1},
Masaaki Nagahara\IEEEauthorrefmark{1},
Kazunori Hayashi\IEEEauthorrefmark{1}, and
Yutaka Yamamoto\IEEEauthorrefmark{1}}
\IEEEauthorblockA{\IEEEauthorrefmark{1}Graduate School of Informatics, Kyoto University, Kyoto, Japan}
}

\maketitle

\begin{abstract}
In this article, we consider the problem of loop-back interference
suppression for
orthogonal frequency division multiplexing (OFDM) signals
in amplify-and-forward single-frequency full-duplex relay stations.
The loop-back interference makes the system a closed-loop system,
and hence it is important not only to suppress the interference but also
to stabilize the system.
For this purpose, we propose 
sampled-data $H^{\infty}$ design of digital filters
that ensure the stability of the system
and suppress the continuous-time effect of interference
at the same time.
Simulation results are shown to illustrate the effectiveness of the proposed method.
\end{abstract}

%
\IEEEpeerreviewmaketitle

%
%
\section{Introduction}
\label{sec:intro}

In wireless communications, relay stations are used to relay radio signals between radio stations that
cannot directly communicate with each other due to the signal attenuation.
On the other hand, it is important to efficiently utilize the scarce bandwidth due to the limitation of frequency resources.
For this purpose, \textit{single-frequency network} is expected in which signals with the same carrier frequency are transmitted through communication network.
Then, a problem of \textit{loop-back interference} caused by coupling waves arises in a 
full-duplex relay station in a single-frequency network \cite{Jain}.

\begin{figure}[t]
		\centering
		\scalebox{0.5}{\includegraphics{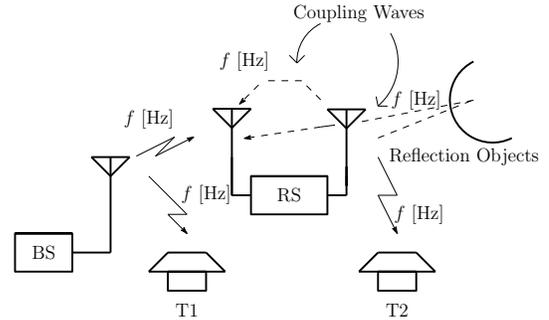}}
		\caption{Loop-back interference}
		\label{coupling}
	\end{figure}

Fig.~\ref{coupling} illustrates loop-back interference by coupling waves.
In this figure, radio signals with carrier frequency $f$~[Hz] are transmitted
from the base station (denoted by BS).
One terminal (denoted by T1) directly receives the signal from the base station,
but the other terminal (denoted by T2) is so far from the base station that they cannot communicate directly.
Therefore, a relay station (denoted by RS) is attached between them to relay radio signals.
Then, radio signals with the same carrier frequency $f$~[Hz] are transmitted from RS to T2,
but also they are fed back to the receiving antenna directly or through reflection objects.
As a result, loop-back interference is caused in the relay station,
which may deteriorate the quality of communication and, even worse,
may destabilize the closed-loop system.

For the problem of loop-back interference,
a simple adaptive filter method with only two antennas has been proposed to cancel the effect of coupling waves in digital domain \cite{sakai2006simple}.
This method does not need multi-antenna array and becomes quite effective when the relay gain is low.
However, it is pointed out that the relay system becomes unstable if the relay gain is very high \cite{Chun}.
Moreover, in theory, if the signals are completely band-limited below the Nyquist frequency,
then discrete-time domain approaches might work well,
but the assumption of perfect band limitedness is hardly satisfied in real communication signals
because real pulse shaping filters do not act perfectly and the nonlinearity
in electric circuits adds frequency components beyond the Nyquist frequency.

To prevent closed-loop stability and consider high-frequency components,
we have proposed a new design method for coupling wave cancelation
based on the sampled-data control theory \cite{SSHRSICE14,SSHR14_01}.
It is introduced to model the transmitted radio signals and coupling waves as continuous-time signals,
and optimize the worst case continuous-time error due to coupling waves by a digital canceler.
This is formulated as a sampled-data $H^{\infty}$ optimal control problem.
Design examples are shown to illustrate the problem methods and evaluate the communication performance.
While our submitted paper \cite{SSHR14_01} shows signal reconstruction performance of the proposed method,
in this manuscript analyzes communication performance.

The remainder of this article is organized as follows.
In Section \ref{sec:models}, we derive a mathematical model of a relay station considered in this study
and propose sampled-data $H^{\infty}$ control for cancelation of loop-back interference.
In Section \ref{sec:sim}, simulation results are shown to illustrate the effectiveness of the proposed method.
In Section \ref{sec:conc}, concluding remarks are offered.

\subsection*{Notation}
Throughout this article, the following notation are used.
We denote by $L^2$ the Lebesgue space consisting of all square integrable real functions 
on $[0, \infty)$ endowed with $L^2$ norm $\|\cdot\|_2$.
The symbol $t$ denotes the argument of time, $s$ the argument of Laplace transform and $z$ the argument of $Z$ transform.
These symbols are used to indicate whether a signal or a system is of continuous-time or discrete-time.
The operator $e^{-Ls}$ with nonnegative real number $L$ denotes the continuous-time delay operator
with delay time $L$.

%
%
\section{Relay Station Model and Filter Design}
\label{sec:models}

In this section, we provide a mathematical model of a relay station with loop-back interference phenomenon
and formulate the cancelation filter design problem.

	\begin{figure}[tb]
	\centering
	\includegraphics[width = \linewidth]{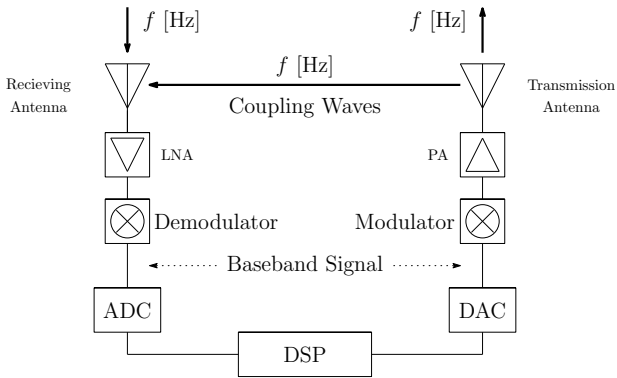}
	\caption{Relay Station}
	\label{Relay Station}
	\end{figure}

There are mainly two types of relaying protocols in wireless communication \cite{Nabar04}.
The first type simply amplifies received digital symbols and transmits them and
is named as amplify-and-forward (AF) protocol.
The other type is called as decode-and-forward (DF) protocol, in which the received signals are demodulated and decoded before transmission.
In the DF protocol, by decoding there does not exist a risk to be unstable.
However, the AF protocol requires quite lower implementation complexity than the DF protocol \cite{Nabar04}.
Throughout this study it is assumed that relay stations operate with the AF protocol.

Fig.~\ref{Relay Station} depicts a typical AF single-frequency full-duplex relay station implemented with a digital canceler.
A radio wave with carrier frequency $f$~[Hz] from a base station is accepted at the receiving antenna and amplified by the low noise amplifier (LNA).
Then, the received signal is demodulated to a baseband signal by the demodulator,
and converted to a digital signal by the analog-to-digital converter (ADC).
The obtained digital signal is then processed and the effect of loop-back interference mentioned in the previous section is suppressed by the digital signal processor (DSP),
which is converted to an analog signal by the digital-to-analog converter (DAC).
Finally, the analog signal is modulated to a radio wave with carrier frequency $f$~[Hz] and transmitted to terminals with which we communicate by the transmission antenna.
Inevitably, this signal is also fed back to the receiving antenna of the relay station.
This is called coupling wave and causes loop-back interference,
which deteriorates the communication quality.

	\begin{figure}[t]
	\includegraphics[width = 0.7\linewidth]{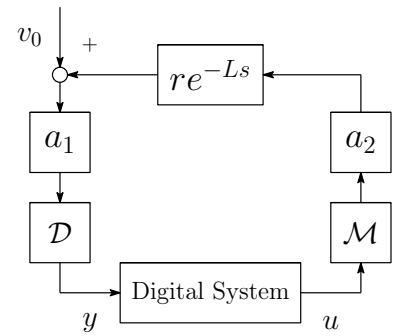}
	\centering
	\caption{Simple Block Diagram of Relay Station}
	\label{Relay Station2}
	\end{figure}

Fig.~\ref{Relay Station2} shows a simplified block diagram of the relay station.
In Fig.~\ref{Relay Station2},
$v_0$ is the transmitted radio wave from the base station.
LNA and PA in Fig.~\ref{Relay Station} are modeled as static gains,
$a_1$ and $a_2$, respectively.
The modulator is denoted by ${\mathcal M}$ and the demodulator by ${\mathcal D}$.
It is assumed that the coupling wave channel is a flat fading channel, that is, all frequency components of a signal through this channel experience the same magnitude fading.
Then the channel can be treated as an all-pass system.
In this study, we adopt a delay system, $re^{-Ls}$,
as a channel model, where $r>0$ is the attenuation rate and $L>0$ is a delay time.
The block named ``Digital System'' includes ADC, DSP, and ADC in Fig.~\ref{Relay Station2}.

In order to realize a single-frequency network, it is considered to use 
orthogonal frequency division multiplexing (OFDM)
with a quadrature amplitude modulation (QAM).
OFDM with QAM has a high equalization performance by the block transmission 
with a cyclic prefix and a good tolerance to inter-symbol interference from multipath environment.
Moreover, this system is easily implemented in a digital signal processor \cite{Goldsmith05}.
QAM transforms a transmission signal into two orthogonal carrier waves,
that is a sine wave and a cosine wave.

\begin{figure}[t]
	\includegraphics[width = 0.9\linewidth]{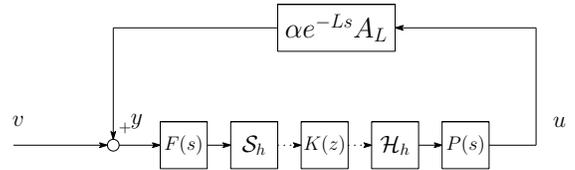}
	\centering
	\caption{Feedback Canceler}
	\label{Feedback Canceler}
	\end{figure}

After some transforms of the block diagram, an equivalent model is gotten as Fig.~\ref{Feedback Canceler} \cite{SSHR14_01}.
In this figure,
$v$ is the received baseband signal
and $\alpha := a_1a_2r$,
\begin{eqnarray}
	A_L := \left[
	\begin{array}{cc}
	\cos 2\pi fL & \sin 2\pi fL \\
	-\sin 2\pi fL & \cos 2\pi fL \\
	\end{array}
	\right].
\end{eqnarray}
$F(s)$ is an anti-aliasing analog filter with an ideal sampler ${\mathcal S_h}$ with sampling period $h>0$,
defined by
\begin{eqnarray}
	\begin{array}{c}
	{\mathcal S}_h : \{ y(t) \} \mapsto \{ y_d[n] \} : y_d[n] = y(nh), \\
     n = 0,1,2,\ldots. \\
	\end{array}
\end{eqnarray}
$K(z)$ denotes a digital filter, which we design for loop-back interference cancelation,
a zero-order hold ${\mathcal H}_h$, is defined by
\begin{eqnarray}
	\begin{array}{c}
	{\mathcal H}_h : \{u_d[n]\} \mapsto \{u(t)\}:u(t)=u_d[n], \\
	t \in [nh, (n+1)h), n=0,1,2,\ldots,\\
	\end{array}
\end{eqnarray}
and a post analog low-pass filter denoted by $P(s)$ model which is a digital-to-analog converter in Fig~\ref{Relay Station}.
It is assumed that $F(s)$ and  $P(s)$ are proper, stable and real-rational transfer function matrices.

\begin{figure}[t]
	\includegraphics[width = \linewidth]{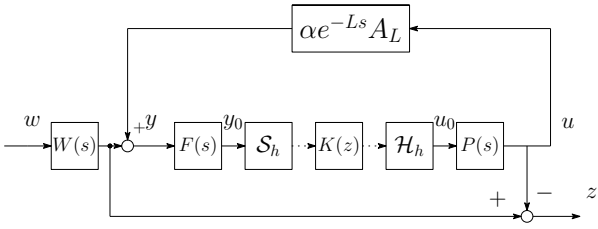}
	\caption{Block Diagram for Feedback Canceler Design}
	\label{Feedback Canceler2}
	\end{figure}

For the obtained system described as Fig.~\ref{Feedback Canceler}, we find the digital controller,
$K(z)$, that stabilizes the feedback system and also minimize the effect of self-interference.
To obtain a reasonable solution, we restrict the input continuous-time signal $v$ to the following set
\begin{eqnarray}
WL^2 := \{v=Ww:w\in L^2, \|w\|_2 = 1\},
\end{eqnarray}
where $W$ is a continuous-time LTI system with real-rational stable, and strictly proper transfer function $W(s)$.
It is notable that OFDM signals are elements of this set.
After that, Fig.~\ref{Feedback Canceler2} is obtained,
with an error signal $z=v-u$.

Let $T_{zw}$ be the system from $w$ to $z$.
Here we formulate our design problem to find a digital controller $K(z)$ that minimizes $\|T_{zw}\|_{\infty}$.
This is a standard sampled-data $H^{\infty}$ control problem, and can be efficiently solved via fast sample fast hold (FSFH) approximation \cite{Chen95}.
Note that if there exists a controller $K(z)$ that minimizes $\|T_{zw}\|_{\infty}$,
then the feedback system shown by Fig.~\ref{Feedback Canceler} is stable and the effect of loop-back interference $z=v-u$ is bounded by the $H^{\infty}$ norm\cite{SSHR14_01}.

%
%
\section{Simulations}
\label{sec:sim}
In the section, simulation results are shown to illustrate the effectiveness of the proposed method.

The bit error rate (BER) is used as a measure of the performance of digital communications.
The rate is defined as the number of bit errors divided by the total number 
of transferred bits over a communication channel
that generates attenuation, noise, interference, and so on.
In addition, the ratio of the energy per bit to the noise power spectral density,
denoted by $E_b/N_0$,
is preferred to indicate the transmission environment performance.
The quantity corresponds to signal-to-noise ratio (SNR) in analog communication.
We show curves of BER vs $E_b/N_0$ and BER vs the relay gain $a_2$
(see Fig.~\ref{Relay Station2})
in order to assess the effectiveness of designed filters.

\begin{figure}[t]
	\includegraphics[width = 0.9\linewidth]{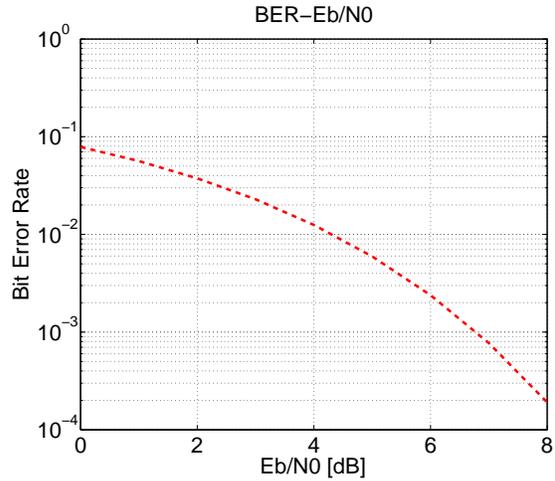}
	\caption{Ideal BER vs $E_b/N_0$ [dB] in probability theory}
	\label{ideal}
	\end{figure}

It is assumed that the sampling period $h$ is normalized to 1,
the normalized carrier frequency $f$ is 10000~[Hz],
the attenuation rate of the coupling wave channel $r=0.15$,
the time delay $L=1$,
and low noise amplifier and the anti-alias analog filter as $a_1=1$ and $F(s) = I$.
The post filter $P(s)$ is modeled by
\begin{eqnarray}
	P(s) = \dfrac{1}{0.001s+1}I.
\end{eqnarray}
With these parameters, we compute the $H^{\infty}$-optimal nominal controller $K(z)$
by FSFH with discretization number $N=16$.
OFDM signals are generated with random binary phase shift keying (BPSK) data of amplitude 1.
The data length is 64 and the guard interval length is 16, namely, 
the block length becomes 80.
It is assumed that the noise is white Gaussian and added at the receiving antenna.
Fig.~\ref{ideal} shows the ideal curve of BER vs $E_b/N_0$
from the receiving antenna to the transmission antenna
without interference nor channel distortion
on the additive white gaussian noise channel.
In this section, we calculate BER with
the input signal to the receiving antenna and the filtered signal by the designed filter, which is denoted by $u$ in Fig.~\ref{Feedback Canceler}.

\begin{figure}[t]
	\includegraphics[width = 0.9\linewidth]{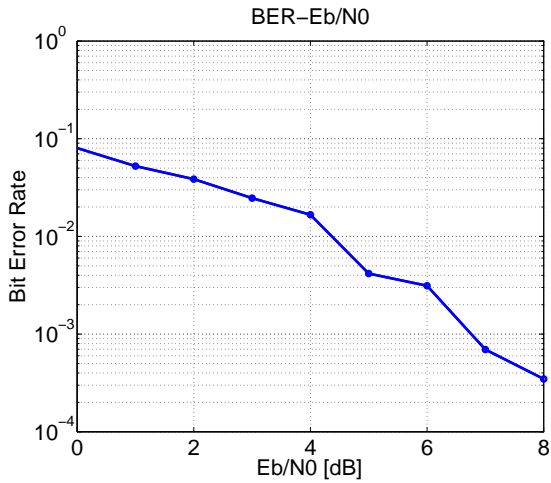}
	\caption{BER vs $E_b/N_0$ [dB] at $a_2 = 2000$ with squared pulse}
	\label{01}
	\end{figure}

\begin{figure}[t]
	\includegraphics[width = 0.9\linewidth]{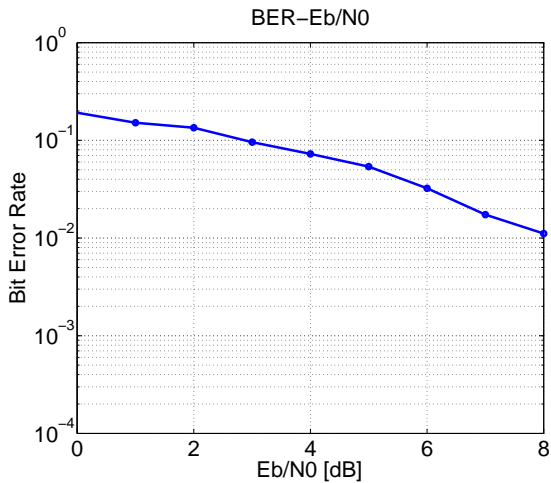}
	\caption{BER vs $E_b/N_0$ [dB] at $a_2 = 2000$ with root-raised-cosine pulse}
	\label{02}
	\end{figure}

Here, $a_2$ is set at $2000$, or, 66 [dB].
First, we employ
a pulse-shaping filter of a sinc type of frequency characteristic
with the symbol period $2h$.
A pulse passed through the filter is described as a squared pulse
since the input signals are impulse ones in the communication considered here.
The frequency characteristic of input signals is modeled by
the following real-rational transfer function
\begin{eqnarray}
W(s) = \dfrac{1}{2s+1}I.
\end{eqnarray}
Fig.~\ref{01} shows the BER vs $E_b/N_0$ using the squared pulse.
Next, we consider the situation where a root-raised-cosine filter is put at the base station and terminals.
Then, the pulse is described as a root-raised-cosine pulse \cite{Goldsmith05}.
The symbol period is assumed to be $3h$,
the roll-off factor is fixed to 0.2,
and the frequency characteristic $W(s)$ is set as
\begin{eqnarray}
W(s) = \dfrac{1}{(2s+1)^4}I.
\end{eqnarray}
Fig.~\ref{02} shows the curve of BER vs $E_b/N_0$ using the root-raised-cosine pulse.
In the simulations, the symbol number is $64 \times 150 = 9600$.

From Fig.~\ref{01}, it can be seen that
with the squared pulse the designed filter can suppress the effect of loop-back interference almost completely.
The curve has lower values than the ideal curve at several points,
but it is inferred that if we simulate more times, the averaged value converges to an upper value.

From Fig.~\ref{02}, if we use
the root-raised-cosine pulse-shaping filter,
the designed filter has lower performance compared to above.
It seems that it happens because the root-raised-cosine pulse has lower high-frequency components
than the squared pulse and then the designed filter becomes more conservative.

\begin{figure}[t]
	\includegraphics[width = 0.9\linewidth]{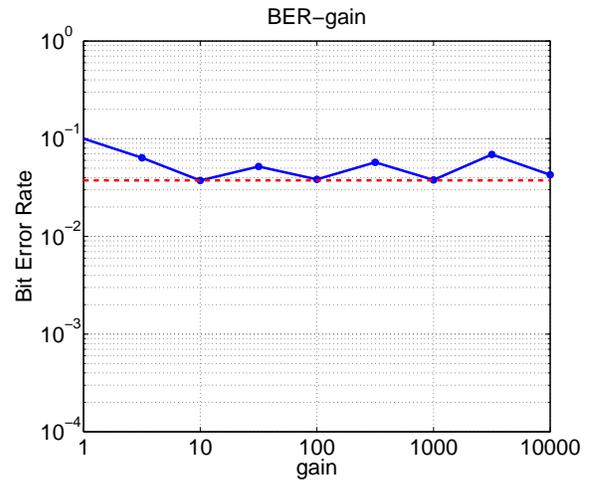}
	\caption{BER vs $a_2$ at $E_b/N_0 = 2$ [dB] with squared pulse: ideal value (dash-dot), loop-back-interference cancelation situation value (solid-line)}
	\label{03}
	\end{figure}

\begin{figure}[t]
	\includegraphics[width = 0.9\linewidth]{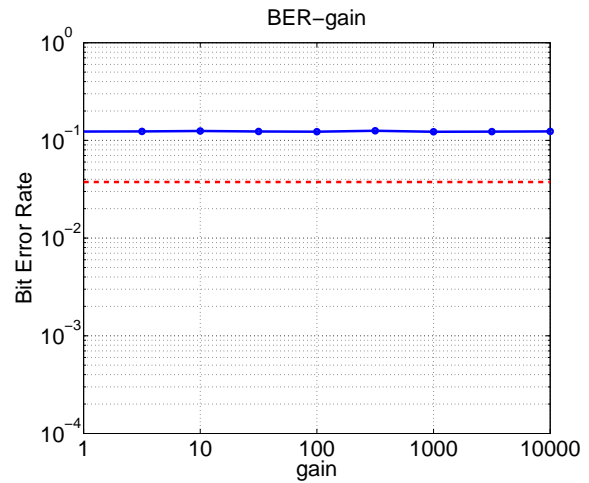}
	\caption{BER vs $a_2$ at $E_b/N_0 = 2$ [dB] with root-raised-cosine pulse: ideal value (dash-dot), loop-back-interference cancelation situation value (solid-line)}
	\label{04}
	\end{figure}

Finally, we check the effect of perturbation in the relay gain $a_2$.
Fig.~\ref{03} depicts the curve of BER vs the gain $a_2$ with a squared pulse 
when $E_b/N_0$ is 2 [dB]
and Fig.~\ref{04} is the curve with a root-raised-cosine pulse.
The dash-dot line shows the ideal values without interference and
the solid line shows the simulated values with the loop-back interference cancelation.
The number of symbols in the simulation is $64 \times 900 = 57600$.
From this figure, if the stability is preserved then BER does not depend 
on the relay gain $a_2$ with the proposed canceler.

%
%
\section{Conclusion}
\label{sec:conc}
In this paper, we have proposed filter design for loop-back interference cancelation
in AF single-frequency full-duplex relay stations based on the sampled-data $H^{\infty}$ control theory.
Simulation results have been shown to illustrate the effect of the proposed cancelers to communication performance.

\section*{Acknowledgement}
This research is supported in part by the JSPS Grant-in-Aid for Scientific Research (B) No.~24360163
and (C) No.~24560543, Grant-in-Aid for Scientific Research on Innovative Areas No.~26120521,
and an Okawa Foundation Research Grant.

\bibliographystyle{bst/IEEEtran}
\bibliography{sshrrefs}

\end{document}